\documentclass[11pt]{article}
\usepackage{amsmath,color}
\usepackage{amssymb}
\usepackage{amsfonts}
\usepackage{graphicx}
\usepackage{color}

\definecolor{darkgreen}{rgb}{0,0.35,0}

\numberwithin{equation}{section}

\begin{document}

\title{\textbf{Stringy horizons and generalized FZZ duality in perturbation theory}}
\author{Gaston Giribet}
\maketitle

\begin{center}

\smallskip
\smallskip
\centerline{Martin Fisher School of Physics, Brandeis University}
\centerline{{\it Waltham, Massachusetts 02453, USA.}}

\medskip
\centerline{Departamento de F\'{\i}sica, Universidad de Buenos Aires
FCEN-UBA and IFIBA-CONICET}
\centerline{{\it Ciudad Universitaria, Pabell\'{o}n I, 1428, Buenos Aires, Argentina.}}

\end{center}

\bigskip

\bigskip

\bigskip





\begin{abstract}
We study scattering amplitudes in two-dimensional string theory on a black hole bakground. We start with a simple derivation of the Fateev-Zamolodchikov-Zamolodchikov (FZZ) duality, which associates correlation functions of the sine-Liouville integrable model on the Riemann sphere to tree-level string amplitudes on the Euclidean two-dimensional black hole. This derivation of FZZ duality is based on perturbation theory, and it relies on a trick originally due to Fateev, which involves duality relations between different Selberg type integrals. This enables us to rewrite the correlation functions of sine-Liouville theory in terms of a special set of correlators in the gauged Wess-Zumino-Witten (WZW) theory, and use this to perform further consistency checks of the recently conjectured Generalized FZZ (GFZZ) duality. In particular, we prove that $n$-point correlation functions in sine-Liouville theory involving $n-2$ winding modes actually coincide with the correlation functions in the $SL(2,\mathbb{R})/U(1)$ gauged WZW model that include $n-2$ oscillator operators of the type described by Giveon, Itzhaki and Kutasov in reference \cite{GIK2}. This proves the GFZZ duality for the case of tree level maximally winding violating $n$-point amplitudes with arbitrary $n$. We also comment on the connection between GFZZ and other marginal deformations previously considered in the literature. 

\end{abstract}

\newpage

\section{Introduction} 

Fateev-Zamolodchikov-Zamolodchikov (FZZ) duality \cite{FZZ, KKK, HS} relates the $n$-point correlation functions of the sine-Liouville integrable model to $n$-point string amplitudes in the Euclidean two-dimensional black hole \cite{Jerusalem, Mandal, Witten}. The latter theory being described by the $SL(2,\mathbb{R})/U(1)$ Wess-Zumino-Witten (WZW) model \cite{Witten}, FZZ duality turns out to be a duality between two seemingly distinct non-rational conformal field theories (CFTs). This is an important result in string theory as it is at the very root of the construction of the black hole matrix model \cite{KKK}. Besides, it has also been important to investigate other string theory scenarios, such as AdS$_3 \times N$ spacetimes \cite{GK} and time-dependent backgrounds \cite{HT}.

FZZ duality has its origin in mirror symmetry \cite{HK}. In fact, the ${\mathcal N}=2$ supersymmetric extension of the duality relates the super-Liouville theory and the Kazama-Suzuki model associated to the $SL(2,\mathbb{R})/U(1)$ coset. In \cite{Maldacena}, Maldacena has shown precisely how the bosonic FZZ duality follows from the supersymmetric duality through a Goodard-Kent-Olive quotient. It follows from this that FZZ can be thought of as a sort of T-duality, which relates two non-compact target spaces of quite different aspects \cite{Seiberg}. The duality is surprising for several reasons; in particular, because it associates two models with different topologies and totally different target space interpretations: While the Euclidean black hole, corresponding to the so-called cigar manifold has topology $\mathbb{R}^2$, the sine-Liouville model is defined on the cylinder $\mathbb{R} \times S$. 

When thought of as a string $\sigma $-model, sine-Liouville has the interpretation of a non-homogeneous tachyon condensate in flat space and in presence of a linear dilaton, while the string theory on the black hole describes a curved spacetime, also in presence of a dilaton configuration that takes the linear form in an appropriate gauge. The FZZ duality can then be interpreted as the equivalence between two different ways of deforming the linear dilaton background, either with the tachyon marginal operator or with the graviton marginal operator \cite{MMP}. Following this interpretation, in a recent paper \cite{GIK2} Giveon, Ithzaki, and Kutasov proposed a generalization of FZZ duality which associates to a larger set of oscillator type primary operators of the WZW theory a set of exponential vertex operators in sine-Liouville theory. This Generalized FZZ conjecture (GFZZ) is supported by the fact that the two kind of operators, even when the corresponding wave functions exhibit different behaviors, actually share the same quantum numbers. In \cite{Giribet3}, a check of GFZZ at the level of the interacting theory was performed by explicitly computing 3-point correlation functions. It was shown there that the duality also holds when vertex interactions are included. The main goal of this paper is to extend the computation of \cite{Giribet3} to the case of $n$-point functions with arbitrary $n$. Unlike the case of the $3$-point functions, for which the result can be explicitly computed in terms of special functions, an analogous expression for the $n$-point correlators with $n>3$ is not at hand. This means that in order to extend the result of \cite{Giribet3} to the $n$-point functions for arbitrary $n$ we have to be more creative: To circumvent such obstruction, we will resort to integral realizations of the $n$-point correlation functions on the Riemann sphere and use it to show the equivalence among different observables.

The paper is organized as follows: In section 2, we will discuss the original FZZ duality. Resorting to the integral representation in terms of Selberg-type integrals, we will show how the $n$-point correlation functions in the sine-Liouville theory can be expressed as $n$-point functions in the WZW model that describes the string $\sigma $-model on the two-dimensional black hole. An intermediate step will be reducing the correlators of the sine-Liouville theory to those of Liouville theory, and then use the $H_3^+$-Liouville correspondence to relate them with the WZW observables. This will lead us to a succinct derivation of the FZZ duality for the sphere topology\footnote{A proof of FZZ correspondence for all genus has been given in \cite{HS} using the path integral approach.}. In section 3, we will use the techniques and results of section 2 to investigate the recently proposed GFFZ duality \cite{GIK2}. We will explicitly show that maximally winding violating $n$-point amplitudes in sine-Liouville theory coincides with the correlation functions in the $SL(2,\mathbb{R})/U(1)$ WZW model that include $n-2$ oscillator operators of the type described in \cite{GIK2}. We will also discuss GFZZ duality in relation to other marginal deformations of the linear dilaton background.

\section{FZZ duality}

\subsection{Step 1: Correlation functions in sine-Liouville theory}

The sine-Liouville field theory is defined by the action
\begin{equation}
S_{\text{sine-L}}[\lambda ]=\frac{1}{2\pi }\int d^{2}z\left( \partial \phi \bar{%
\partial }\phi +\partial X\bar{\partial }X+\frac{R {Q} \phi }{\sqrt{%
2}}+4\pi \lambda \ e^{\frac{1}{\sqrt{2}{Q}}\phi }\cos \Big( \sqrt{\frac{k}{2}}%
\tilde{X}\Big) \right)   \label{sL}
\end{equation}
where $\phi (z,\bar{z}) $ is a scalar field that takes values in $\mathbb{R}$ and presents a background charge ${Q} = -1/\sqrt{k-2}$, while the field $X (z,\bar{z})$ lives on a circle of radius $\sqrt{k}$ and represents an angular direction; thus, the target space interpretation of this model describes an Euclidean space with topology $\mathbb{R}\times S^1$. The field $\tilde X (z,\bar{z})= X_L (z) - X_R (\bar{z})$ that appears in the interaction term of (\ref{sL}) is the T-dual of $X (z,\bar{z})= X_L (z) + X_R (\bar{z})$.

Action (\ref{sL}) defines a non-compact conformal field theory with central charge 
\begin{equation}
c=2+\frac{6}{k-2} \label{Yc}
\end{equation}
whose interpretation within the context of string theory is that of a two-dimensional linear dilaton background in presence of a non-homogeneous tachyon condensate. The tachyon potential resembles a Liouville wall that prevents the strings from exploring the strongly coupled region, with the string coupling being 
\begin{equation}
g_s(\phi )=e^{-\sqrt{\frac{2}{k-2}}\phi }. 
\end{equation}

One of the reasons why the duality between the sine-Liouville theory (\ref{sL}) and the Euclidean black hole $\sigma$-model is remarkable is that, as mentioned, it relates two models with different topology; while sine-Liouville is defined on the cylinder, the Euclidean black hole resembles the geometry of a semi-infinite cigar and thus has topology $\mathbb{R}^2$. This is related to the fact that in each of these models the mechanism by means of which the strings are prevented from going to the strong coupling region is different: Both models are asymptotically --i.e. at large $\phi $ in (\ref{sL})-- equivalent to a cylinder of radius $\sqrt{k}$. In the case of the Euclidean black hole, the asymptotic cylindrical region is far from the horizon, with the compact direction of the cylinder corresponding to the Euclidean time. In both models the string coupling constant vanishes in the asymptotic cylinder. In sine-Liouville theory, the Liouville type wall is what prevents the strings from entering into the strong coupling zone, while in the black hole such region is not accessible simply because it lies behind the horizon and thus does not belong to the Euclidean manifold\footnote{See \cite{GIK} for an interesting discussion on the relevance of the inner black hole region for describing the high-energy string scattering and how it relates to FZZ.}.

The vertex operators that create primary states in the theory (\ref{sL}) are given by
\begin{eqnarray}
V^{\ p,\omega }_{j}(z,\bar{z}) = e^{\sqrt{\frac{2}{k-2}}j\phi(z,\bar{z})} e^{i\sqrt{\frac{2}{k}}p X(z,\bar{z})+i\sqrt{{2k}}\omega \tilde{X}(z,\bar{z})} \ , \label{Y23}
\end{eqnarray}
where the quantum numbers $j, p$, and $\omega $ represent the momentum  in $\sqrt{k}$ units along the longitudinal direction $\phi $, the momentum along the compact direction $X$, and the winding number around the latter, respectively. Notice that, unlike $p$, the winding number $\omega $ is not a conserved quantity in this model due to the presence of the interaction term. It is customary to define the variables 
\begin{equation}
m = \frac{1}{2} (p + k\omega ) \ , \ \ \ \bar{m} = \frac{1}{2} (p - k\omega ) \ , \label{Y24}
\end{equation}
which facilitate the comparison with the WZW observables.

Another notable difference between sine-Liouville theory and the Euclidean black hole $\sigma$-model is the mechanisms by means of which each CFT allows for the violation of the winding number conservation: In sine-Liouville theory the winding number is not conserved because of the explicit dependence on $\tilde{X}$ in the potential. This is usually rephrased as the winding number being broken by a condensate of wound strings. In contrast, in the cigar CFT the winding number is not conserved simply because the wound strings can in principle slip through the tip of the cigar. Explicit computation of WZW correlators actually confirms this picture of the winding number violation in both theories \cite{GN, MO2, FH}.

In terms of the variables (\ref{Y24}), the vertex operators (\ref{Y23}) read
\begin{eqnarray}
V_{j,m,\bar{m}}(z,\bar{z}) = e^{\sqrt{\frac{2}{k-2}}j\phi(z,\bar{z})} e^{i\sqrt{\frac{2}{k}}mX(z)+i\sqrt{\frac{2}{k}}\bar{m}\bar{X}(\bar{z})}\label{Y25}
\end{eqnarray}
where now we are denoting $X(z)=X_L(z)$ and $\bar{X}(\bar{z})=X_R(\bar{z})$ for short. The conformal dimension of these vertex operators is given by
\begin{equation}
\Delta = -\frac{j(j+1)}{k-2} + \frac{m^2}{k} \ , \ \ \ \bar\Delta = -\frac{j(j+1)}{k-2} + \frac{\bar{m}^2}{k} \ . \label{Y26}
\end{equation}

The interaction term in (\ref{sL}) can be written as a particular integrated vertex operator (\ref{Y25}); namely
\begin{equation}
{2\lambda } \int d^2z \ \cos \Big(\sqrt{\frac{k}{2}} \tilde X (z,\bar{z})\Big) = {\lambda } \int d^2z  \ V_{1-\frac{k}{2},\frac{k}{2},\frac{k}{2}} (z,\bar{z}) + {\lambda } \int d^2z \ V_{1-\frac{k}{2},\frac{k}{2},-\frac{k}{2}} (z,\bar{z}) . \label{Y27}
\end{equation}
This leads to define the two screening charges 
\begin{equation}
Q_{\pm} = \lambda \int d^2z \ V_{1-\frac{k}{2},\frac{k}{2},\pm \frac{k}{2}} (z,\bar{z}), \label{Y28}
\end{equation} 
which can be used to give an integral realization of correlation functions. 

Correlation functions in sine-Liouville model are defined as follows
\begin{equation}
\mathcal{A} _{\text{sine-L}}= \Big\langle  \prod_{i=1}^n \ :V_{j_i , m_i, \bar{m}_i} (z_i, \bar{z}_i): \ \Big\rangle_{\text{sine-L}} = \int {\mathcal D}\phi {\mathcal D}X e^{-S_{\text{sine-L}}[\lambda ]}\ \prod_{i=1}^n V_{j_i , m_i, \bar{m}_i} (z_i, \bar{z}_i) 
\end{equation}
and standard CFT techniques permit to write the residues of these correlation functions as\footnote{We will abuse the notation and not make distinction between the correlation functions and the associate residues. There is a relative factor $\sim \Gamma (1+s)\Gamma (-s)$ when relating both quantities. This factor comes from the integration of the zero-mode $\phi_0 $, and it produces a divergence in the case of resonant correlators $s\in \mathbb{Z}_{\geq 0}$. Such divergence is associated to the non-compactness of the target space. Analytic extension of the expressions above to values $s\in \mathbb{C}$ is required in order to gather the most general correlators.}
\begin{eqnarray}
\mathcal{A}_{\text{sine-L}} = \frac{\lambda^{\tilde{s} + s}}{\tilde{s}!s!} &&
\int \prod_{l=1}^{s} d^2u_l \int \prod_{r=1}^{\tilde{s}} d^2v_r 
\int {\mathcal D}\phi {\mathcal D}X  e^{-S_{\text{sine-L}}[\lambda =0]}
\prod_{i=1}^n V_{j_i , m_i, \bar{m}_i} (z_i, \bar{z}_i) 
\times \nonumber \\
&&\ \ \prod_{l=1}^{s} V_{1-\frac{k}{2},\frac{k}{2},\frac{k}{2}} (u_l,\bar{u}_l)
\prod_{r=1}^{\tilde{s}} V_{1-\frac{k}{2},\frac{k}{2},-\frac{k}{2}} (v_r,\bar{v}_r) \label{Y210}
\end{eqnarray}
with
\begin{eqnarray}
\sum_{i=1}^{n}j_i + 1 - \frac{k-2}{2} (\tilde{s} + s) = 0 \ , \label{JK}
\end{eqnarray}
which now reduces to a $(n+s+\tilde{s})$-point correlator of the free theory $\lambda =0$. Using (\ref{Y210}), one can resort to the operator product expansion (OPE) in the free theory, namely
\begin{equation}
V_{j_i,m_i,\bar{m}_i} (z_i, \bar{z}_i) V_{j_j,m_j,\bar{m}_j} (z_j, \bar{z}_j) \simeq (z_i - z_j)^{-\frac{2}{k-2}j_ij_j+\frac{2}{k}m_im_j} (\bar{z}_i - \bar{z}_j)^{-\frac{2}{k-2}j_ij_j+\frac{2}{k}\bar{m}_i\bar{m}_j}
\end{equation}
and, after Wick contraction, write
\begin{eqnarray}
\mathcal{A}_{\text{sine-L}} &=& \frac{\lambda^{\tilde{s} + s}}{\tilde{s}!s!}
\prod_{i<j}^n (z_i - z_j)^{-\frac{2}{k-2}j_ij_j+\frac{2}{k}m_im_j} (\bar{z}_i - \bar{z}_j)^{-\frac{2}{k-2}j_ij_j+\frac{2}{k}\bar{m}_i\bar{m}_j}  \nonumber \\
&& \int \prod_{l=1}^{s} d^2u_l \int \prod_{r=1}^{\tilde{s}} d^2v_r
\prod_{l=1}^{s}\prod_{r=1}^{\tilde{s}} |u_ l - v_r|^{2-2k}
\prod_{l<l'}^{s} |u_l - u_{l'}|^{2}
\prod_{r<r'}^{\tilde{s}} |v_r - v_{r'}|^{2}
\nonumber \\
&&\prod_{i=1}^n \prod_{l=1}^{s} (z_i - u_l)^{j_i+m_i} (\bar{z}_i - \bar{u}_l)^{j_i-\bar{m}_i}
\prod_{i=1}^n \prod_{r=1}^{\tilde{s}} (z_i - v_r)^{j_i-m_i} (\bar{z}_i - \bar{v}_r)^{j_i+\bar{m}_i} 
. \label{Integral}
\end{eqnarray}

Condition (\ref{JK}) comes from the integration over the zero mode of $\phi $. In addition, the following conditions come from the integration over the zero mode of $X$,
\begin{eqnarray}
\sum_{i=1}^{n}m_i + \frac{k}{2} s - \frac{k}{2} \tilde{s}= 0 \ , \ \ \ \ \sum_{i=1}^{n}\bar{m}_i - \frac{k}{2} s + \frac{k}{2} \tilde{s}= 0 .
\end{eqnarray}
These conditions translate into the conservation equations
\begin{equation}
\Delta p \equiv \sum_{i=1}^n p_i = 0 \ , \ \ \ \ \Delta \omega \equiv\sum_{i=1}^n \omega_i = \tilde{s} -s
\end{equation}
from which we observe that the different relative amounts of operators of the type $Q_-$ and $Q_+$ used to screen the background charge in (\ref{Integral}) translate into different values of the total winding number $\Delta \omega $; in particular, $\tilde{s}=s \leftrightarrow \Delta\omega=0$. In other words, computing correlation functions with $\Delta \omega $ amounts to introduce the operators
\begin{equation}
(Q_-)^{s+\Delta \omega } (Q_+)^{{s}}  ,
\end{equation} 
in the free theory correlation functions and then perform the Coulomb gas calculation. These correlators were explicitly computed in reference \cite{FH} for the case $n=3$ with $\tilde{s}-s=1$, where it was shown that it yields an expression in terms of the $\Upsilon $-functions that appear in the Dorn-Otto-Zamolodchikov-Zamolodchikov (DOZZ) formula \cite{elDO, ZZ}. We will see below how this computation can be done in an alternative way.

\subsection{Multiple Selberg integrals}

Now, let us explain how to deal with the integrals of the type (\ref{Integral}). These are defined integrating over $\mathbb{C}$ with the measure $ d^2u_l = (i/2)du_l d\bar{u}_l$ (with, say, $u_l = x_l + i y_l$, $\bar{u}_l = {x}_l - i {y}_l$), and can be computed by an appropriate contour prescription\footnote{To integrate, it is convenient to Wick rotate $x_l\to ix_l$ and introduce a deformation parameter in $|u_l |^2=-x_l ^2+y_l ^2+i\epsilon $ in order to avoid the poles that emerge at $x_l =\pm y_l $. The way to proceed is first to define coordinates $x^{\pm }_l = \pm  x_l + y_l $ and integrate over $x^-_l$ while keeping the $x^+_l$ fixed.}. In the particular case $n=3$, this can be explicitly computed and it yields the Dotsenko-Fateev formula \cite{DF}. Here, however, we are interested in the case $n\geq 3$ with $\lambda_i \neq 0$. Fortunately, we will not need to solve the integral (\ref{Integral}) explicitly, but only exploiting some of its functional properties. 

Let us define  
\begin{eqnarray}
I^N_{s,\nu }(y_1,...y_{N};\delta_1,...\ \delta_{N};\lambda_1,...\ \lambda_{N}) \equiv \frac{1}{s!} \int \prod_{r=1}^s \frac{d^2 v_r}{\pi } 
\prod_{r=1}^s\prod_{a=1}^N (v_r - y_a)^{\delta_a - \lambda_a} (\bar{v}_r - \bar{y}_a)^{\delta_a + \lambda_a} \prod_{r<r'}^s |v_r - v_{r'}|^{2\nu } \nonumber
\end{eqnarray}
A remarkable property that this integral obeys for generic $N$ is the following \cite{BF}
\begin{eqnarray}
I^N_{s,1}(y_1,...y_{N};\delta_1,...\ \delta_{N};\lambda_1,...\ \lambda_{N}) &=& I^N_{N-s-1,1}(y_1,...y_{N};-1-\delta_1,...\ -1-\delta_{N};-\lambda_1,...\ -\lambda_{N})
\nonumber \\
&&\prod_{a<b}^N (y_a- y_b)^{1+\delta_a+\delta_b-\lambda_a-\lambda_b} (\bar{y}_a- \bar{y}_b)^{1+\delta_a+\delta_b+\lambda_a+\lambda_b}  \nonumber \\
&&  \frac{\Gamma(-s-\sum_{a=1}^n(\delta_a-\lambda_a))}{\Gamma(1+s+\sum_{a=1}^n(\delta_a+\lambda_a))} \prod_{a=1}^{N} \frac{\Gamma(1+\delta_a-\lambda_a)}{\Gamma(-\delta_a-\lambda_a)} 
\label{fateevidentity}
\end{eqnarray}
which generalizes the well known formula for the integral $I^2_{1,\nu }(y_1,y_{2};\delta_1,\delta_{2};0,0)$, i.e. with $s=1$, $N=2$, namely
\begin{equation}
\int d^2v \ |v-y_1|^{2\delta _1} |v-y_2|^{2\delta _2} = \pi |y_1 - y_2|^{2+\delta_1 +\delta_2} \frac{\Gamma(-1-\delta_1 -\delta_2) \Gamma(1+\delta_1) \Gamma(1+\delta_2) }{ \Gamma(2+\delta_1 +\delta_2) \Gamma(-\delta_1) \Gamma(-\delta_2)} \ ,
\end{equation}
or its chiral analogue with $\lambda_i \neq 0 $. Formula (\ref{fateevidentity}) has appeared in the literature \cite{BF} in different contexts; it has been used to explicitly compute correlation functions \cite{FL}, as well as to provide an alternative succinct derivation of the Dotsenko-Fateev formula \cite{FL2}. Here, following an idea of Fateev \cite{Fateev}, we will use functional relation (\ref{fateevidentity}) to translate sine-Liouville correlators into Liouville theory correlators. The first step to do so is to integrate out in (\ref{Y210}) the $\tilde{s}$ vertices inserted at the points $v_1, v_2, v_3, ... \ v_{\tilde{s}}$, and then transform that integral into the following
\begin{eqnarray}
\mathcal{A}_{\text{sine-L}} = \frac{\lambda^{2s +\Delta \omega }\pi^{s+\Delta\omega }}{s! } 
\Bigg( \frac{\Gamma(2-k)}{\Gamma(k-1)}\Bigg)^{s}
\prod_{i=1}^{n} \frac{\Gamma(1+j_i-m_i)}{\Gamma(-j_i-\bar{m}_i)} \ \ \ \ \ \ \ \ \ \ \ \ \ \ \ \ \  \ \ \ \ \ \ \ \ \ \ \ \ \ \ 
 \nonumber \\
\ \prod_{i<j}^n (z_i - z_j)^{\eta^+_{ij}}
(\bar{z}_i - \bar{z}_j)^{\bar{\eta}^-_{ij}}\  \int \prod_{l=1}^{s} d^2u_l \prod_{i=1}^n \prod_{l=1}^{s} |z_i - u_l|^{4j_i-2(k-2)} 
\prod_{l<l'}^{s} |u_l - u_{l'}|^{-4(k-2)} \ \ \ \ \ \ \ \ \ \
 \nonumber \\
I^{n+s}_{n-\Delta\omega -1,1}(z_1,...z_{n},u_1,...u_{s};-1-j_1+\frac{k}{2}\omega_1,...-1-j_n+\frac{k}{2}\omega_n,k-2,...k-2;-\frac{p_1}{2},...-\frac{p_n}{2},0...0) \nonumber
\end{eqnarray}
with
\begin{equation}
\eta^{\pm }_{ij}= -\frac{2}{k-2}\left( j_i+1-\frac{k}{2}\right) \left( j_j+1-\frac{k}{2}\right) +\frac{2}{k}\left( m_i\mp \frac{k}{2} \right) \left( m_j\mp \frac{k}{2} \right) \ .
\end{equation}


This expression can be conveniently rearranged as follows
\begin{eqnarray}
\mathcal{A}_{\text{sine-L}} = 
\frac{\pi^{2-n}c_k^{\Delta \omega } }{ (n-\Delta \omega -2)!} \ \prod_{i=1}^{n} \frac{\Gamma(1+j_i-m_i)}{\Gamma(-j_i-\bar{m}_i)}
\prod_{i<j}^n (z_i - z_j)^{\frac{2}{k} (m_i -\frac{k}{2})(m_j -\frac{k}{2})} (\bar{z}_i - \bar{z}_j)^{\frac{2}{k} (\bar{m}_i +\frac{k}{2})(\bar{m}_j +\frac{k}{2})} 
\nonumber \\
 \int \prod_{r=1}^{n-\Delta\omega-2} d^2y_r \ \prod_{r=1}^{n-\Delta\omega-2} \prod_{r=1}^{n} (z_i - y_r)^{m_i-\frac{k}{2}} (\bar{z}_i - \bar{y}_r)^{-\bar{m}_i-\frac{k}{2}}\prod_{r<r'}^{n-\Delta\omega-2} |y_r - y_{r'}|^{k}
\ \ \ \ \ \ \ \ \label{lobo} \\
\  \tilde{\mu }^s\ \prod_{i<j}^n |z_i - z_j |^{-4\alpha_i\alpha_j}  I^{2n-\Delta\omega-2}_{s,-2b^{-2}} (z_1, ... z_n,y_1,...y_{n-\Delta\omega-2}; -\frac{2\alpha_1}{b}, ...-\frac{2\alpha_n}{b},\frac{1}{b^2}, ...\frac{1}{b^2}; 0,...0 ) \nonumber 
\end{eqnarray}
where, in addition, we have used projective symmetry to fix one of the integration variables at infinity, say $y_{n-\Delta\omega -1} = \infty$. For convenience, we have also defined the quantities
\begin{equation}
N^{j_i}_{m_i,\bar{m}_i}=\frac{\Gamma(1+j_i-m_i)}{\Gamma(-j_i-\bar{m}_i)} \ , \ \ \ \ \ \ \ b^2=\frac{1}{k-2} \ , \ \ \ \ \ \ \ \alpha_i = b(b^{-2}/2-j_i ) \ , \label{alfa}
\end{equation}
together with
\begin{equation}
\tilde{\mu }= \pi^2 \lambda^2\frac{ \Gamma(-b^{-2})}{\Gamma(1+b^{-2})}\ , \ \ \ \ \ \ \ c_k=\pi^2\lambda . \label{Z223}
\end{equation}

In the next subsection we will see how (\ref{lobo}) can actually be rewritten in terms of correlation functions of Liouville field theory. Later, in subsection 2.4, we will show that (\ref{lobo}) can also be written as correlators in the $SL(2,\mathbb{R})/U(1)$ WZW theory. 

\subsection{Step 2: Reduction to Liouville correlation functions}

The remarkable fact, early observed by Fateev \cite{Fateev}, is that the last line of (\ref{lobo}) can be identified as a $(2n-\Delta \omega -2)$-point function in Liouville theory. In fact, Liouville $N$-point correlation functions, which are defined by 
\begin{equation}
\mathcal{A} _{\text{L}}= \Big\langle  \prod_{a=1}^N \ :e^{\sqrt{2}\alpha_a\phi (z_a, \bar{z}_a)}: \ \Big\rangle _{\text{L}} = \int {\mathcal D}\phi e^{-S_{\text{L}}[\mu ]}\ \prod_{a=1}^N  e^{\sqrt{2}\alpha_a\phi (z_a, \bar{z}_a)}\label{corrL}
\end{equation}
with the Liouville action
\begin{equation}
S_{\text{L}}[\mu ]=\frac{1}{2\pi }\int d^{2}z\left( \partial \phi \bar{%
\partial }\phi +\frac{RQ_L \phi }{2\sqrt{
2}}+2\pi \mu \ e^{\sqrt{2}b\phi} \right) \ \ \ , \ \ Q_L =b+\frac{1}{b},   \label{L}
\end{equation}
admit an integral representation of the form
\begin{equation}
\mathcal{A}_{\text{L}}= \tilde{\mu }^s\ \prod_{a<b}^N |z_a - z_b|^{-4\alpha_a\alpha_b}\  I_{s,-2b^{-2}}^N (z_1, ... \ ,  z_N; -\frac{2\alpha_1}{b}, ...\ ,  -\frac{2\alpha_N}{b} ; 0,... \ , 0)\label{Y226}
\end{equation}
with
\begin{equation}
s=b^2+1-b\sum_{a=1}^N\alpha_a , \label{scre}
\end{equation}
and where $\tilde{\mu }$ is related to the constant that appears in the action (\ref{L}) through \cite{ZZ}
\begin{equation}
\Bigg( \pi \tilde{\mu } \frac{\Gamma(b^{-2})}{\Gamma(1-b^{-2})}\Bigg)^{\frac{b}{2}}=\Bigg( \pi \mu \frac{\Gamma(b^2)}{\Gamma(1-b^2)}\Bigg)^{\frac{b^{-1}}{2}}.\label{Y228}
\end{equation}

Expression (\ref{Y226}) follows from the existence of a second marginal operator in the theory, which is not the one that appears in action (\ref{L}); namely $\exp (\sqrt{2}\phi /b)$. This reflects the self-duality of Liouville theory under $(\mu , b)\leftrightarrow (\tilde{\mu },1/b)$, cf. (\ref{Y48}) below.

Then, we observe that the last line of (\ref{lobo}) corresponds to (\ref{Y226}) with $N=2n-\Delta \omega -2$, and $\alpha _i =-1/(2b)$ for $i=n+1, ...\ 2n-\Delta \omega -2$. Therefore, one observes that the sine-Liouville correlation functions can be written in terms of Liouville correlation functions by the following convolution formula\footnote{Hereafter we omit the normal ordering symbols :: for short. }
\begin{eqnarray}
\mathcal{A}_{\text{sine-L}} &=& 
\frac{\pi^{2-n}c_k^{\Delta \omega } }{ (n-\Delta \omega -2)!} \ \prod_{i=1}^{n} N^{j_i}_{m_i,\bar{m}_i}\
\int \prod_{r=1}^{n-\Delta\omega-2} d^2y_r
\Big\langle  \prod_{i=1}^n \  e^{\sqrt{2}\phi (z_i, \bar{z}_i)}  \ \prod_{r=1}^{n-\Delta\omega -2} \  e^{-\frac{1}{\sqrt{2}b}\phi (y_r, \bar{y}_r)}  \ \Big\rangle_{\text{L}}
\nonumber \\
&&
\ \times \ \Big\langle  \prod_{i=1}^n \  e^{i\sqrt{\frac{2}{k}}(\frac{k}{2}-m_i)\chi (z_i)+ i\sqrt{\frac{2}{k}}(\frac{k}{2}+\bar{m}_i) \bar{\chi}(\bar{z}_i)}  \
\prod_{r=1}^{n-\Delta \omega -2 }  e^{i\sqrt{\frac{k}{2}}\chi (y_r)+ i\sqrt{\frac{k}{2}} \bar{\chi}(\bar{y}_r)}  \ \Big\rangle_{\text{free}}\label{resultado1}
\end{eqnarray}
where we have also defined
\begin{eqnarray}
&&\Big\langle  \prod_{i=1}^n \  e^{i\sqrt{\frac{2}{k}}(\frac{k}{2}-m_i)\chi (z_i)+ i\sqrt{\frac{2}{k}}(\frac{k}{2}+\bar{m}_i) \bar{\chi}(\bar{z}_i)}  \
\prod_{r=1}^{n-\Delta \omega -2 }  e^{i\sqrt{\frac{k}{2}}\chi (y_r)+ i\sqrt{\frac{k}{2}} \bar{\chi}(\bar{y}_r)}  \ \Big\rangle_{\text{free}} = \nonumber \\
&&
\ \ \ \ \ \ \prod_{i<j}^n (z_i - z_j)^{\frac{2}{k}(\frac{k}{2}-m_i)(\frac{k}{2}-m_j)} (\bar{z}_i - \bar{z}_j)^{\frac{2}{k}(\frac{k}{2}+\bar{m}_i)(\frac{k}{2}+\bar{m}_j)}
\nonumber \\
&&
\ \ \ \ \ \prod_{r=1}^{n-\Delta\omega-2} \prod_{r=1}^{n} (z_i - y_r)^{m_i-\frac{k}{2}} (\bar{z}_i - \bar{y}_r)^{-\bar{m}_i-\frac{k}{2}}
\prod_{r<r'}^{n-\Delta\omega-2} |y_r - y_{r'}|^{k} \ ,
\end{eqnarray}
which, actually, can be interpreted as the correlation function of a free scalar $\chi $ with background charge\footnote{The total central charge of Liouville theory cupled with the $U(1)$ field $\chi $ is $c=1+6Q_L^2+1-6|Q_{\chi }|^2 = 2+6/(k-2)$, which matches (\ref{Yc}).} $Q_{\chi }=i\sqrt{k}$ \cite{Giribet, GL, Giribet2}. 

Formula (\ref{resultado1}) expresses that sine-Liouville can be represented by the product of Liouville theory and a $c<1$ CFT. More precisely, sine-Liouville $n$-point correlation function that violates the total winding number in $\Delta \omega $ units can be written in terms of a Liouville $(2n-\Delta\omega -2)$-point correlation function\footnote{In particular, this provides an alternative way of computing the sine-Liouville 3-point function calculated in \cite{FH}, as formula (\ref{resultado1}) translates this problem into that of evaluating the Liouville DOZZ structure constants.} that includes $n-\Delta\omega -2$ states with momentum $\alpha _{i}=-1/(2b)$, with $i=n+1, ... \ 2n-\Delta\omega -2$. Such particular value of the Liouville momentum $\alpha $ corresponds to a degenerate representation of Virasoro algebra; that is, to a highest-weight representation that contains null states in the Virasoro Verma module. This is crucial to make connection with the WZW correlators. In fact, in the following subsection we will review how one can also associate to correlator (\ref{resultado1}) a correlator in the WZW theory. This follows from the so-called $H_3^+$-Liouville correspondence \cite{Stoyanovsky, RT, HS2, Ribault}, which maps the Belavin-Polyakov-Zamolodchikov (BPZ) equation obeyed by the Liouville correlators that include degenerate representations with the Knizhnik-Zamolodchikov (KZ) equation obeyed by WZW correlators. However, before moving to that, let us review some properties of Liouville correlation functions that will be needed later: Let us begin by recalling that, apart from the integral representation (\ref{Y226})-(\ref{Y228}), Liouville $N$-point correlation functions admits the alternative integral representation
\begin{equation}
\mathcal{A}_{\text{L}}= {\mu }^{\hat{s}} \ \prod_{a<b}^N |z_a - z_b|^{-4\alpha_a\alpha_b}\  I_{\hat{s},-2b^{2}}^N (z_1, \ ... \ ,  z_N; -2{\alpha_1}{b}, \ ... \ ,  -2{\alpha_N}{b} ; 0, \ ... \ ,  0) \label{Y48}
\end{equation}
with 
\begin{equation}
\hat{s} = b^{-2}s = -b^{-1}\sum_{a=1}^N \alpha_a +b^{-2}+1.\label{elAsterisco}
\end{equation}
This expression is the $ b \leftrightarrow 1/b$ dual to (\ref{Y226})-(\ref{Y228}). Representation (\ref{Y48})-(\ref{elAsterisco}) is probably more natural as it follows from perturbation theory in the theory (\ref{L}), \cite{elGL}. 

Also, it will be useful to be reminded of the reflection properties of Liouville correlation functions: Under the momentum reflection $\alpha \to \alpha^*=Q_L-\alpha$, which leaves the conformal dimension of the exponential operators invariant, Liouville correlation functions satisfy the following property \cite{ZZ}
\begin{equation}
\Big\langle  \prod_{a=1}^N \  e^{\sqrt{2}\alpha_a\phi (z_a, \bar{z}_a)}  \ \Big\rangle _{\text{L}} = R(\alpha_1 ) \ \Big\langle   e^{\sqrt{2}\alpha_1^*\phi (z_1, \bar{z}_1)} \ \prod_{a=2}^N \  e^{\sqrt{2}\alpha_a\phi (z_a, \bar{z}_a)}  \ \Big\rangle _{\text{L}} \label{Y412}
\end{equation}
where
\begin{equation}
R(\alpha ) = -\Bigg( \pi \mu \frac{\Gamma(b^2)}{\Gamma(1-b^2)} \Bigg)^{\frac{Q-2\alpha }{b}} \frac{\Gamma(2\alpha b-b^2 )\Gamma(2\alpha b^{-1}-b^{-2})}{(Q_L-2\alpha )^2\Gamma(1-2\alpha b+b^2 )\Gamma(1-2\alpha b^{-1}+b^{-2})}
\end{equation}
is the reflection coefficient, which is closely related\footnote{The reflection coefficient differs from the overall factor of the 2-point function by a factor $\pi^{-1}(Q_L-2\alpha )$, \cite{Zamolodchikov}.} to the 2-point function \cite{ZZ}.

\subsection{Step 3: Connection with the WZW correlation functions and FZZ}

Now, let us explain how sine-Liouville $n$-point correlation functions (\ref{resultado1}) can also be written as $n$-point correlation functions in the WZW theory. This follows from the main result of reference \cite{Ribault}, which actually proves that the right hand side of (\ref{resultado1}) is exactly the expression of a $SL(2,\mathbb{R})/U(1)$ WZW correlators\footnote{Our notation relates to that in \cite{Ribault} by a Weyl reflection of the $SL(2,\mathbb{R})$ isospin variable, $j\to -1-j$, together with a $\mathbb{Z}_2$ transformation $\bar{m}\to -\bar{m}$; the latter maps the $SL(2,\mathbb{R})$ highest-weight discrete representations ${\mathcal D}_j^+$ into the lowest-weight discrete representations ${\mathcal D}_j^-$ and vice versa. There is also an extra global factor $\pi^{1-n} $ that can be absorbed in the normalization of the vertices and of the path integral. Notice that here we managed to give a precise value for the coefficient $c_k$ in terms of $\lambda $ and, consequently, of $\mu $, while in \cite{Ribault} it remains unspecified.}. The result of \cite{Ribault} was generalizing the so-called $H^3_+$-Liouville correspondence \cite{Stoyanovsky, RT} to the case in which the winding number is not necessarily conserved. Therefore, putting (\ref{resultado1}) and the result\footnote{See formula (3.29) therein.} of \cite{Ribault} together, we conclude
\begin{equation}
{\mathcal A}_{\text{sine-L}}\equiv \Big\langle \prod_{i=1}^{n} \  V_{j_i,m_i,\bar{m}_i} (z_i,\bar{z}_i) \ \Big\rangle_{\text{sine-L}} = \Big\langle \prod_{i=1}^{n} \  V_{j_i,m_i,\bar{m}_i} (z_i,\bar{z}_i) \ \Big\rangle_{SL(2,\mathbb{R})/U(1)} \label{FZZduality}
\end{equation}
where on the right hand side we have the correlator in the gauged $SL(2,\mathbb{R})/U(1)$ WZW theory. Expression (\ref{FZZduality}) is the FZZ duality. 

The $SL(2,\mathbb{R})/U(1)$ WZW correlator in (\ref{FZZduality}) involves primary operators\footnote{We decided to employ the same notation both for the sine-Liouville vertex operators (\ref{Y25}) and for these operators of the $SL(2,\mathbb{R})/U(1)$ WZW theory. We do this to emphasize the interpretation of FZZ duality as the equivalence between two different ways of deforming the same theory by adding different marginal operators.} $V_{j_i,m_i,\bar{m}_i} $ that are constructed from the primary operators $\Phi_{j_i,m_i,\bar{m}_i}^{\omega_i } $ of the ungauged $SL(2,\mathbb{R})$ theory. The latter operators create Kac-Moody primary states $\vert j,m \rangle\otimes \vert j,\bar{m} \rangle$ of Hermitian unitary representations $j$ of $SL(2,\mathbb{R})$ in the spectral flow sector $\omega $ \cite{MO, GN}. The coset projection amounts to impose the condition\footnote{Notice that in this convention there is a minus sign in the label $\bar{m}_i$ with respect to \cite{Ribault}. This convention is consistent with the winding number around the cigar to be $\omega = (m-\bar{m})/k$.} $m_i -\bar{m}_i-k\omega_i =0 $, which is consistent with (\ref{Y24}). This permits to associate the winding number around the direction $X$ of the coset to the spectral flow parameter $\omega $ in the $SL(2,\mathbb{R})$ theory. In the sector $\Delta \omega =0$, the correlators in the coset theory relates to those in the $SL(2,\mathbb{R})$ theory by a prefactor\footnote{See references \cite{MO, GN, Giribet2} for the details of the construction of the spectral sectors of the $SL(2,\mathbb{R})$ WZW in terms of the product $SL(2,\mathbb{R})/U(1)\ \times U(1)$.}
\begin{equation}
\prod_{i<j}^{n} (z_i - z_j)^{\frac{2}{k}m_im_j}(\bar{z}_i - \bar{z}_j)^{\frac{2}{k}\bar{m}_i\bar{m}_j}
\end{equation}

The WZW theory for the coset $SL(2,\mathbb{R})/U(1)$ has central charge
\begin{equation}
c=\frac{3k}{k-2}-1, \label{yaestocomolollamamos}
\end{equation}
where $k$ is the WZW level; the term $-1$ comes from the $U(1)$ gauging. Notice this matches (\ref{Yc}). In the semiclassical limit, $k\to \infty $, one obtains $c=2$ as expected.

So far, we did not need to compute the WZW correlation functions explicitly. The trick to avoid this was resorting to the $H_3^+$-Liouville correspondence. However, in the next section we will need to compute a new set of WZW correlation functions, and therefore we will need to introduce a convenient framework to do so. A useful technique is the Wakimoto free field approach --see \cite{BB, GN} and references therein--. In terms of the Wakimoto free fields, the action of the $SL(2,\mathbb{R})$ WZW model reads
\begin{equation}
S_{\text{WZW}}[\mu ]=\frac{1}{2\pi }\int d^{2}z\left( \partial \phi \bar{%
\partial }\phi +\beta \bar{\partial}\gamma + \bar{\beta} {\partial}\bar{\gamma } -\frac{R\phi }{\sqrt{%
2(k-2)}}+2\pi \mu \beta \bar{\beta } \ e^{-\sqrt{\frac{2}{k-2}}\phi }\right)   \label{sWZW}
\end{equation}
which involves a $\beta$-$\gamma$ system and a real scalar field $\phi (z,\bar{z}) $. The latter represents the longitudinal direction of the cigar, corresponding to the radial coordinate in the black hole geometry. The dilaton varies linearly along $\phi $ and the string coupling vanishes in the asymptotic region, far from the horizon. The horizon is located at the tip of the cigar, i.e. at $\phi=0$. The value of the dilaton at the horizon can be adjusted by shifting the zero mode of $\phi $, which changes the value of $\mu $. This means that the constant\footnote{Notice that here we are denoting the black hole mass by $\mu $. This is because when relating Liouville and WZW correlators the black hole mass turns out to match the Knizhnik-Polyakov-Zamolodchikov (KPZ) scaling coefficient of Liouville theory --often referred to as the Liouville cosmological constant--. In the case of sine-Liouville theory the relation is given by (\ref{Z223}) and (\ref{Y228}). For instance, in the case of the theory with $c=26$, for which $k=9/4$, the KPZ scaling for the sine-Liouville spherical partition function is $\lambda^8$, as discussed in \cite{KKK}.} $\mu $, which physically corresponds to the mass of the two-dimensional black hole, is linked to the string coupling.

The black hole $\sigma $-model is not given by the $SL(2,\mathbb{R})$ theory (\ref{sWZW}) but by the gauged $SL(2,\mathbb{R})/U(1)$ model. The latter can be constructed from the former by the following recipe \cite{BB, BK, DVV}: In addition to (\ref{sWZW}), one considers a free boson $X$,
\begin{equation}
S_X = \frac{1}{2\pi }\int d^2z \ \partial X \bar{\partial } X 
\end{equation}
and a copy of the $c=-2$ $B$-$C$ ghost CFT 
\begin{equation}
S_{BC} = \frac{1}{2\pi }\int d^2z ( \bar{C}\partial \bar{B} + {C}\bar{\partial} {B})  \ ;
\end{equation}
then, one considers the BRST charge
\begin{eqnarray}
Q_{U(1)}=\oint dz \ C(z) \Bigg( J^3(z)-i\sqrt{\frac{k}{2}}\partial X(z)\Bigg) , \label{Q241}
\end{eqnarray}
with $J^3(z)$ being the local $SL(2,\mathbb{R})$ current that generates the $\hat{u}(1)$ affine symmetry of the $U(1)$ direction of the coset. The $SL(2,\mathbb{R})$ currents $J^{3}(z)$, $J^{-}(z)$, $J^{+}(z)$ have the following OPE with operators
\begin{eqnarray}
J^3 (z) \Phi_{j,m,\bar{m}} (w,\bar{w}) &\simeq & \frac{m}{(z-w)} \Phi_{j,m,\bar{m}} (w,\bar{w}) + ... \\
J^{\pm } (z) \Phi_{j,m,\bar{m}} (w,\bar{w}) &\simeq & \frac{\pm j-m}{(z-w)} \Phi_{j,m\pm 1,\bar{m}} (w,\bar{w}) + ... 
\end{eqnarray}
where the ellipses stand for regular terms; and analogously for the anti-holomorphic components. This means that one can define the operators $V_{j,m,\bar{m}} $ of the $SL(2,\mathbb{R})/U(1)$ coset as follows
\begin{eqnarray}
V_{j,m,\bar{m}} (z,\bar{z}) = \Phi_{j,m,\bar{m}} (z,\bar{z}) \times e^{i\sqrt{\frac{2}{k}}mX(z)-i\sqrt{\frac{2}{k}}\bar{m}\bar{X}(\bar{z})} \label{esos}
\end{eqnarray}
which, indeed, have conformal dimension (\ref{Y26}) and are annihilated by (\ref{Q241}). These are the operators appearing on the right hand side of (\ref{FZZduality}). In the next section, we will use this free field representation to compute correlation functions in the coset theory and investigate the GFZZ duality.

\section{GFZZ duality}

\subsection{Black hole curvature operator}

An efficient way to compute correlation functions in the non-compact WZW theory is to consider the free theory $S_{\text{WZW}}[\mu =0]$ perturbed by the dimension-(1,1) operator
\begin{equation}
\tilde{V}_{1,1} (z,\bar{z}) =  \beta (z)\bar{\beta }(\bar{z}) \ e^{-\sqrt{\frac{2}{k-2}}\phi (z,\bar{z}) } \label{Easterisco}
\end{equation}
and resort to the free field representation \cite{BB, GN}. A remark that will be relevant below is that operator (\ref{Easterisco}) can be expressed as
\begin{eqnarray}
\tilde{V}_{1,1} (z,\bar{z}) = J^+(z) \bar{J}^+(\bar{z}) \ \Phi_{-1,-1,-1}(z,\bar{z}) \label{Y23bisa}  
\end{eqnarray}
where $ J^+(z)$ is the local Kac-Moody current associated to the upper triangular element of $SL(2,\mathbb{R})$. It is possible to verify that (\ref{Y23bisa}) is actually an operator of the $SL(2,\mathbb{R})/U(1)$ coset CFT, as $J^3(z)J^+(0)\simeq J^+(0)/z$ + regular terms, while $J^3(z) \Phi_{-1,-1,-1}(0)\simeq - \Phi_{-1,-1,-1}(0)/z$ + regular terms. A way to verify that (\ref{Easterisco}) is actually realized by (\ref{Y23bisa}) is to use that in terms of the Wakimoto free fields the vertex operators of the $SL(2,\mathbb{R})$ theory are given by
\begin{eqnarray}
\Phi_{j,m,\bar{m}} (z,\bar{z}) = \gamma^{j-m}(z)  \ e^{\sqrt{\frac{2}{k-2}}j\phi(z)} \times \text{h.c.}\label{cosos}
\end{eqnarray}
where h.c. stands for the anti-holomorphic counterpart that depends on $\bar{m}$, and that the current $ J^+(z)$ is simply given by the field $\beta (z) $. Then, the regular part of OPE between these fields yields (\ref{Easterisco}). This is the first member of a family of operators with which we will be concerned here. (\ref{Easterisco}) is not of the form (\ref{cosos}), it rather belongs to a different representation. In fact, apart from (\ref{esos}), there are other operators in the coset which are relevant for our discussion; namely 
\begin{eqnarray}
\tilde{V}_{\ell , \bar{\ell}} (z,\bar{z}) = \   (J^+(z))^{\ell } (\bar{J}^+(\bar{z}))^{\bar{\ell } } \ \Phi_{-\frac{\ell+\bar{\ell }}{2},-\frac{\ell+\bar{\ell }}{2},-\frac{\ell+\bar{\ell }}{2}}(z,\bar{z}) \label{Y23bis}  
\end{eqnarray}
which are also of the coset --operator (\ref{Easterisco}) corresponds to the particular case $\ell =\bar{\ell }=1$--. In terms of Wakimoto free fields, operator (\ref{Y23bis}) reads
\begin{equation}
\tilde{V}_{\ell,\bar{\ell }} (z,\bar{z}) = {\mathcal N}_{\ell ,\bar{\ell }}\ (\beta (z))^{\ell} (\bar{\beta }(\bar{z}))^{\bar{\ell }} \ e^{-\sqrt{\frac{2}{k-2}}\frac{(\ell + \bar{\ell })}{2}\phi (z,\bar{z})}. \label{tildeO1}
\end{equation}
with ${\mathcal N}_{\ell ,\bar{\ell }}$ being a normalization. These $\beta $-dependent primary vertex operators have been considered in the literature before --see for instance \cite{BK, GN, Lopez, Giribet2}-- and play an important role in the formulation of the GFZZ duality of \cite{GIK2}. In the following subsections we will study the properties of these operators.

\subsection{Oscillator states and black hole amplitudes}

As mentioned, the FZZ duality can be thought of as a duality between the graviton-like operator $\tilde{V}_{1,1}$, which controls the curvature of the black hole $\sigma $-model, and the tachyon-like operators
\begin{equation}
V_{\frac{k}{2}, \frac{k}{2},\pm \frac{k}{2}} (z,\bar{z}) =  e^{-\sqrt{\frac{k-2}{2}}\phi (z,\bar{z})+i\sqrt{\frac{k}{2}}X(z)\pm i\sqrt{\frac{k}{2}}\bar{X}(\bar{z})}, \label{O2}
\end{equation}
appearing in sine-Liouville action. In fact, both operators have the same quantum numbers. In a  recent paper \cite{GIK2}, Giveon, Itzhaki and Kutasov proposed that the the relation between (\ref{Easterisco}) and (\ref{O2}) might be regarded as the simplest example of a larger duality between operators in the Euclidean black hole $\sigma$-model and tachyon-type operators in sine-Liouville theory. More precisely, this GFZZ duality states the correspondence between oscillator-type operators of the form (\ref{Y23bis}) and the tachyon-like operators\footnote{Our notation relates to that of \cite{GIK2} by $j\to -1-j$. The origin of the duality between these operators of the coset theory is the equivalence between states of the $SL(2,\mathbb{R})$ discrete representation ${\mathcal D}_{j}^{\pm }$ of the spectral flow sector $\omega $ and those of the discrete representation ${\mathcal D}_{-k/2-j}^{\mp }$ of the spectral flow sector $\omega \pm 1$ \cite{GIK2}; this is similar to the relation between conjugate representations of the vertex operators \cite{GN}.}
\begin{equation}
V_{\frac{\ell+\bar{\ell}-k}{2},\frac{k-\ell+\bar{\ell}}{2},\frac{\bar{\ell }-\ell -k}{2}} (z,\bar{z}) =  e^{-\sqrt{\frac{2}{k-2}}(\frac{k-\ell-\bar{\ell}}{2})\phi (z,\bar{z})+i\sqrt{\frac{2}{k}}(\frac{k-\ell+\bar{\ell}}{2})X(z)- i\sqrt{\frac{2}{k}}(\frac{k+\ell-\bar{\ell}}{2})\bar{X}(\bar{z})}. \label{tildeO2}
\end{equation} 

The conformal dimension of both operators is indeed the same; it is given by
\begin{equation}
\Delta = \ell - \frac{(\ell + \bar{\ell }) (\ell + \bar{\ell }-2)}{4(k-2)} \ , \ \ \ \bar{\Delta }= \bar{\ell } - \frac{(\ell + \bar{\ell }) (\ell + \bar{\ell }-2)}{4(k-2)} , 
\end{equation}
which in the case $\ell=\bar{\ell}=1$ corresponds to the marginal deformations (\ref{Easterisco}) with $\Delta =\bar{\Delta} =1$. 

In \cite{Giribet3} this GFZZ duality has been checked at the level of the interacting theory by explicitly computing $3$-point functions that involve both (\ref{tildeO1}) and (\ref{tildeO2}). Here, we aim to go further and perform a consistency check of the duality at the level of $n$-point functions for arbitrary $n$. As we said, this demands to be more creative than in \cite{Giribet3} because, unlike what happens in the case $n=3$, for $n>3$ an expression for the correlation functions in terms of elementary functions is not available. The way of circumventing this obstruction is using the integral representation of $n$-point functions discussed in the previous section. We will focus for simplicity on the case in which the winding number number is violated maximally; that is, on the $n$-point correlation functions that include $n-2$ operators of the type (\ref{tildeO2}) in sine-Liouville theory. We will explicitly show that, indeed, such correlation functions exactly match the $n$-point correlation functions involving the appropriate amount of oscillator type states (\ref{tildeO1}) in the Euclidean black hole $\sigma $-model, in agreement with \cite{GIK2}. More concretely, we want to compute
\begin{equation}
\tilde{\mathcal A}_{\text{WZW}} = \prod^{n-2}_{i=1}  {\mathcal N}_{\ell_i ,\bar{\ell }_i}\ \Big\langle
\prod_{i=1}^{n-2} \  \tilde {V}_{\ell_i, \bar{\ell}_i}(z_i , \bar{z}_i) \
  \  {V}_{j_{n-1},m_{n-1},\bar{m}_{n-1}}(1 , 1) 
\  {V}_{j_{n},m_{n},\bar{m}_{n}}(0 , 0) \    \Big\rangle_{SL(2,\mathbb{R})/U(1)} \label{Y46}
\end{equation}
and show that this exactly agrees with the sine-Liouville correlator that involves $n-2$ operators (\ref{tildeO2}). We will do this in the following subsection.

\subsection{Oscillator states amplitudes}

Let us compute in the coset WZW theory the correlation function (\ref{Y46}), which involves $n-2$ oscillator-type operators $\tilde{V}_{\ell ,\bar{\ell }}$. For our purpose, it is enough to consider the case $\ell_i=\bar{\ell}_i$ which in the sine-Liouville side corresponds to a correlator with $m_{i}-\bar{m}_{i}=k$ for $i\leq n-2$ and with $m_{n-1}=\bar{m}_{n-1}=-m_{n}=-\bar{m}_{n}$, i.e. $\omega _{i}=1$ for $i\leq n-2$, and $\omega _{n-1}=\omega_n=0$. This yields $\Delta \omega =n-2$. Direct computation of the WZW correlator using Wakimoto representation, similar to the one carried out in \cite{Giribet3}, yields the integral representation
\begin{eqnarray}
\tilde{\mathcal{A}}_{\text{WZW}}= {\mu ^{\hat{s}}}   \frac{\Gamma(1+j_{n-1}-m_{n-1})\Gamma(1+j_{n}-m_{n})}{\Gamma(-j_{n-1}-m_{n-1})\Gamma(-j_{n}-m_{n})} \ \prod^{n-2}_{i=1} (-1)^{\ell_{i}} {\mathcal N}_{\ell_i ,{\ell }_i} \ \ \ \ \ \nonumber \\
\ \ \ \ \ \ I_{\hat{s},-2b^2}^{n} (z_1,...z_{n-2},1,0;-2\ell_1 b^2,...-2\ell_{n-2} b^2, 2j_{n-1}b^2-1, 2j_{n}b^2-1; 0, ... 0) \label{MMM}
\end{eqnarray}
with
\begin{equation}
\hat{s}+\sum_{i=1}^{n-2}\ell_i -\sum_{l=n-1}^{n}j_l-1=0.\label{Ufas}
\end{equation}
The $\Gamma $-functions in (\ref{MMM}) come from the multiplicity factor of the Wick contraction of $\beta - \gamma $ fields of the ghost system, which have the free field correlator $\langle \beta (z) \gamma (w) \rangle =1/(z-w)$; see references \cite{BB, GN, Giribet2, Giribet3} for details.
 
Using the integral representation (\ref{Y48}) for the Liouville correlation functions, and comparing (\ref{elAsterisco}) with (\ref{Ufas}), one observes that (\ref{MMM}) can be written in terms of Liouville theory correlator as follows
\begin{eqnarray}
\tilde{\mathcal{A}}_{\text{WZW}}&=& \frac{\Gamma(1+j_{n-1}-m_{n-1})\Gamma(1+j_{n}-m_{n})}{\Gamma(-j_{n-1}-m_{n-1})\Gamma(-j_{n}-m_{n})} \prod_{i=1}^{n-2}(-1)^{\ell _i}{\mathcal N}_{\ell_i ,\ell_i} \nonumber \\
&&\Big\langle \prod_{i=1}^{n-2} \  e^{\sqrt{2}\alpha^*_i \phi(z_i,\bar{z}_i)}  \  e^{\sqrt{2}\alpha_{n-1} \phi(1,1)}  \  e^{\sqrt{2}\alpha_{n} \phi(0,0)}  \ \Big\rangle_{\text{L}}  \label{Y414}
\end{eqnarray}
with
\begin{eqnarray}
\alpha_i &=& b(b^{-2} /2 -j_i ), \ \ \ i=n-1, n \ , \label{mapa22222} \\
\ell_i&=&{k}/{2}+j_i \ , \ \ \ \ \ \ \ \ i=1,2,...\ n-2 \ . \label{mapa}
\end{eqnarray}

Then, considering the correspondence between sine-Liouville operators and Liouville correlators (\ref{resultado1}) in the special case $\Delta\omega =n-2$, using the reflection property (\ref{Y412}) of Liouville correlation functions, using the identity between WZW and Liouville correlators \cite{Ribault} for the case $\Delta \omega = n-2$, and defining the normalization ${\mathcal N}_{\ell_i , \ell_i}$ conveniently in order to absorb factors $R(\alpha _i)$ and $\Gamma (1+\ell_i -k) / \Gamma (k-\ell_i ) $, one finally finds that the $SL(2,\mathbb{R})/U(1)$ WZW correlator
\begin{eqnarray}
\tilde{\mathcal A}_{\text{WZW}} = \Big\langle \prod_{i=1}^{n-2} \   \tilde{V}_{\ell_i , \ell_i} (z_i , \bar{z}_i)   \   V_{j_{n-1},m_{n-1}, \bar{m}_{n-1}} (1,1)   \   V_{j_{n},m_{n}, \bar{m}_{n}} (0,0)   \ \Big\rangle_{SL(2,\mathbb{R})/U(1)} \label{BOBO1}
\end{eqnarray}
is actually equivalent to the sine-Liouville correlator\footnote{Similar computation can be carried out for chiral correlators with $\ell_i \neq \bar{\ell }_i$ provided one considers $\lambda _i \neq 0 $ in (\ref{fateevidentity}).}
\begin{eqnarray}
{\mathcal A}_{\text{sine-L}} = \Big\langle \prod_{i=1}^{n-2} \   {V}_{\ell_i -\frac{k}{2}, \frac{k}{2}, - \frac{k}{2}} (z_i , \bar{z}_i)   \   V_{j_{n-1},m_{n-1}, \bar{m}_{n-1}} (1,1)   \   V_{j_{n},m_{n}, \bar{m}_{n}} (0,0)   \ \Big\rangle_{\text{sine-L}} , \label{BOBO2}
\end{eqnarray}
with the dictionary (\ref{mapa}). This exactly agrees with the GFZZ conjecture of \cite{GIK2}. The matching between correlators (\ref{BOBO1}) and (\ref{BOBO2}) is consistent with --and gives further support to-- the identification between the operators (\ref{tildeO1}) and (\ref{tildeO2}), namely 
\begin{eqnarray}
\tilde{V}_{\ell , \bar{\ell }}  \ \ \ \  \leftrightarrow \ \ \ \ V_{\frac{\ell+\bar{\ell}-k}{2},\frac{k-\ell+\bar{\ell}}{2},\frac{\bar{\ell}-\ell-k}{2}} . \label{ESAAAA}
\end{eqnarray}
This operator map is remarkable for several reasons: 
First, it relates states with different winding number. Secondly, it relates operators whose associated wave functions have quite different support: While the large $\phi $ behavior of one of these operators is 
$\tilde{V}_{\ell , \ell} \sim \exp {(-\ell \sqrt{{2}/({k-2})}\phi )}$, the other goes like ${V}_{\ell -k/2 , k/2 , k/2} \sim \exp {(\sqrt{{2}/({k-2})}(\ell -k/2)\phi )}$. From this, we observe that also the large semiclassical -- large $k$ -- limit of both operators is different; while the former goes like $\tilde{V}_{\ell , \ell} \sim \exp {(-\ell \sqrt{({2}/{k})}\phi )}$ the latter behaves as ${V}_{\ell -k/2 , k/2 , k/2} \sim \exp {(- \sqrt{{k}/{2}}\phi )}$.

\subsection{Other marginal deformations}

Duality (\ref{ESAAAA}) is a generalization of the identification between the marginal operator (\ref{O2}) of sine-Liouville theory and the metric operator (\ref{Easterisco}) of the two-dimensional black hole $\sigma $-model. In particular, (\ref{ESAAAA}) also relates other marginal operators in both theories. For instance, in the WZW theory there exists a second marginal operator\footnote{See \cite{GN} for explicit computations involving this operator.}, whose Wakimoto realization is given by
\begin{equation}
\tilde{V}_{k-2,k-2} (z,\bar{z}) = (\beta (z) )^{k-2} (\bar{\beta }(\bar{z}) )^{k-2} e^{-\sqrt{2(k-2)}\phi (z,\bar{z})} \ ,
\end{equation}
and corresponds to $\ell = \bar{\ell }=k-2$. Therefore, according to the identification (\ref{ESAAAA}), this operator should correspond in the sine-Liouville theory to the dimension-(1,1) operator
\begin{equation}
V_{\frac{k}{2}-2,\frac{k}{2},-\frac{k}{2}} (z,\bar{z})= e^{\Big(\sqrt{\frac{k-2}{2}}-\sqrt{\frac{2}{k-2}}\Big) \phi (z,\bar{z})} e^{i\sqrt{\frac{k}{2}} X(z)-i\sqrt{\frac{k}{2}} \bar{X}(\bar{z})}. \label{B314}
\end{equation}
Then, a natural question arises as to what is the role played by the marginal operator (\ref{B314}) in sine-Liouville theory. It turns out that this operator has already appeared in the literature previously: In \cite{MMP}, Mukherjee, Mukhi and Pakman considered the sine-Liouville theory deformed by the operator (\ref{B314}), and they showed that the resulting theory, being now defined with the two marginal operators (\ref{Y27}) and (\ref{B314}), is actually consistent with the structure of sine-Liouville correlation functions. More importantly, they also showed that the OPE of the two exponential marginal operators actually generates --and demands the presence of-- the black hole deformation operator (\ref{Easterisco}). This peculiar feature is somehow explained by (\ref{ESAAAA}).

Another special case of the GFZZ mapping (\ref{ESAAAA}) is $\ell = \bar{\ell }=0$, which corresponds to the identity operator $\tilde{V}_{0,0} = 1 $ in the WZW theory. It turns out that (\ref{ESAAAA}) maps the identity operator to the dimension-(0,0) operator
\begin{equation}
V_{-\frac{k}{2},\frac{k}{2},-\frac{k}{2}} (z,\bar{z})= e^{-\frac{k}{2(k-2)} \phi(z,\bar{z}) +i\sqrt{\frac{k}{2}}X(z)-i\sqrt{\frac{k}{2}}\bar{X}(\bar{z})} . \label{CI}
\end{equation}
This operator (\ref{CI}) has also appeared in the literature before \cite{FZZ}, and it has been regarded as a {\it conjugate identity operator}, which plays an important role in the computation of winding violating amplitudes \cite{FZZ, GN, MO2, Giribet2, Giribet3}. In \cite{MO2}, operator (\ref{CI}) has been reconsidered in the context of string theory on AdS$_3$, and was identified as the {\it the spectral flow operator}. The existence of this conjugate identity operator was intriguing from the worldsheet theory point of view; however, in the light of (\ref{ESAAAA}) this is nothing but another manifestation of the GFZZ duality.

Further identifications among different operators that appeared in the literature are somehow explained by the GFZZ map (\ref{ESAAAA}). For instance, in the case $\ell = \bar{\ell }= k-1$, (\ref{ESAAAA}) happens to connect different conjugate representations of the identity $\tilde{\mathcal I}_0$, $\tilde{\mathcal I}_-$ considered in reference \cite{GN}, which turn out to be important to compute correlators with winding string states. The identification between operators (\ref{ESAAAA}) permits to see from a different perspective the relation between different conjugate representations of primary operators of arbitrary weight; see for instance the field $\tilde{\Phi }^{\omega =-1}_{j,m=k/2,\bar{m}=k/2}$ of \cite{Lopez}, which corresponds here to the oscillator operator $\tilde{V}_{j+k/2,j+k/2}$. From this point of view, correspondence (\ref{ESAAAA}) appears as a particular case of the map of conjugated representations
\begin{equation}
\tilde{V}_{j,m,\bar{m}}=(\beta(z))^{j+m} (\bar{\beta}(\bar{z}))^{j+\bar{m}} e^{-\sqrt{\frac{2}{k-2}}(j+\frac{k}{2})\phi(z,\bar{z})+i\sqrt{\frac{2}{k}}(m-\frac{k}{2})X(z)+i\sqrt{\frac{2}{k}}(\bar{m}-\frac{k}{2})\bar{X}(\bar{z})}   \  \leftrightarrow \ V_{j,m,\bar{m}} \nonumber
\end{equation} 
proposed in reference \cite{Lopez}.

\section{Conclusions}

We have shown that $n$-point correlation functions in sine-Liouville theory involving $n-2$ winding modes coincide with the correlation functions in the $SL(2,\mathbb{R})/U(1)$ gauged WZW model that involve $n-2$ operators of the oscillator type introduced in reference \cite{GIK2}. Our result is consistent with the Generalized FZZ correspondence, and it extends our previous work \cite{Giribet3} to the case of $n$-point functions.

\[ \]
\subsection*{Acknowledgments}

I am grateful to V. Fateev for having provided me a few years ago with a copy of his unpublished notes \cite{Fateev}. I would also like to thank the members of the Center for Cosmology and Particle Physics (CCPP) of New York University (NYU) for their hospitality during my stay, where this work was finished. This work has been supported by CONICET through the grant PIP 0595/13, and by NSF/CONICET bilateral cooperation program.

\[ \]

\providecommand{\href}[2]{#2}\begingroup\raggedright\endgroup

\begin{thebibliography}{10}





\bibitem{GIK2}
  A.~Giveon, N.~Itzhaki and D.~Kutasov,
  ``Stringy Horizons II,''
  JHEP {\bf 1610}, 157 (2016)
  [arXiv:1603.05822 [hep-th]].


\bibitem{FZZ} V. Fateev, A. Zamolodchikov, and Al. Zamolodchikov,
unpublished.


\bibitem{KKK} 
  V.~Kazakov, I.~K.~Kostov and D.~Kutasov,
  ``A Matrix model for the two-dimensional black hole,''
  Nucl.\ Phys.\ B {\bf 622}, 141 (2002)
  [hep-th/0101011].


\bibitem{HS} 
Y.~Hikida and V.~Schomerus,
  ``The FZZ-Duality Conjecture: A Proof,''
  JHEP {\bf 0903}, 095 (2009)
  [arXiv:0805.3931 [hep-th]].


\bibitem{Jerusalem} 
  S.~Elitzur, A.~Forge and E.~Rabinovici,
  ``Some global aspects of string compactifications,''
  Nucl.\ Phys.\ B {\bf 359}, 581 (1991).


\bibitem{Mandal} 
  G.~Mandal, A.~M.~Sengupta and S.~R.~Wadia,
  ``Classical solutions of two-dimensional string theory,''
  Mod.\ Phys.\ Lett.\ A {\bf 6}, 1685 (1991).

\bibitem{Witten}
  E.~Witten,
  ``On string theory and black holes,''
  Phys.\ Rev.\ D {\bf 44}, 314 (1991).

\bibitem{GK} 
  A.~Giveon and D.~Kutasov,
  ``Notes on AdS(3),''
  Nucl.\ Phys.\ B {\bf 621}, 303 (2002)
  [hep-th/0106004].


\bibitem{HT} 
  Y.~Hikida and T.~Takayanagi,
  ``On solvable time-dependent model and rolling closed string tachyon,''
  Phys.\ Rev.\ D {\bf 70}, 126013 (2004)
  [hep-th/0408124].

	
\bibitem{HK} 
  K.~Hori and A.~Kapustin,
  ``Duality of the fermionic 2-D black hole and N=2 liouville theory as mirror symmetry,''
  JHEP {\bf 0108}, 045 (2001)
  [hep-th/0104202].

\bibitem{Maldacena} 
  J.~M.~Maldacena,
  ``Long strings in two dimensional string theory and non-singlets in the matrix model,''
  JHEP {\bf 0509}, 078 (2005)
  [Int.\ J.\ Geom.\ Meth.\ Mod.\ Phys.\  {\bf 3}, 1 (2006)]
  [hep-th/0503112].

	\bibitem{Seiberg} 
  N.~Seiberg,
  ``Emergent spacetime,''
  hep-th/0601234.

\bibitem{MMP} 
  A.~Mukherjee, S.~Mukhi and A.~Pakman,
  ``FZZ Algebra,''
  JHEP {\bf 0701}, 025 (2007)
  [hep-th/0606037].

\bibitem{Giribet3} 
  G.~Giribet,
  ``Scattering of low lying states in the black hole atmosphere,''
  Phys.\ Rev.\ D {\bf 94}, no. 2, 026008 (2016)
  Addendum: [Phys.\ Rev.\ D {\bf 94}, no. 4, 049902 (2016)]
  [arXiv:1606.06919 [hep-th]].

\bibitem{GIK} 
A.~Giveon, N.~Itzhaki and D.~Kutasov,
  ``Stringy Horizons,''
  JHEP {\bf 1506}, 064 (2015)
  [arXiv:1502.03633 [hep-th]].

\bibitem{GN} 
 G.~Giribet and C.~A.~Nunez,
  ``Correlators in AdS(3) string theory,''
  JHEP {\bf 0106}, 010 (2001)
  [hep-th/0105200].


\bibitem{MO2} 
  J.~M.~Maldacena and H.~Ooguri,
  ``Strings in AdS(3) and the SL(2,R) WZW model. Part 3. Correlation functions,''
  Phys.\ Rev.\ D {\bf 65}, 106006 (2002)
  [hep-th/0111180].



\bibitem{FH}
  T.~Fukuda and K.~Hosomichi,
  ``Three point functions in sine-Liouville theory,''
  JHEP {\bf 0109}, 003 (2001)
  [hep-th/0105217].

\bibitem{elDO}
  H.~Dorn and H.~J.~Otto,
  ``Two and three point functions in Liouville theory,''
  Nucl.\ Phys.\ B {\bf 429}, 375 (1994)
  [hep-th/9403141].
	
\bibitem{ZZ}
  A.~B.~Zamolodchikov and A.~B.~Zamolodchikov,
  ``Structure constants and conformal bootstrap in Liouville field theory,''
  Nucl.\ Phys.\ B {\bf 477}, 577 (1996)
  [hep-th/9506136].


\bibitem{DF}
  V.~S.~Dotsenko and V.~A.~Fateev,
  ``Four Point Correlation Functions and the Operator Algebra in the Two-Dimensional Conformal Invariant Theories with the Central Charge $c < 1$,''
  Nucl.\ Phys.\ B {\bf 251}, 691 (1985).


\bibitem{BF} 
  P.~Baseilhac and V.~A.~Fateev,
  ``Expectation values of local fields for a two-parameter family of integrable models and related perturbed conformal field theories,''
  Nucl.\ Phys.\ B {\bf 532}, 567 (1998)
  [hep-th/9906010].


\bibitem{FL} 
  V.~A.~Fateev and A.~V.~Litvinov,
  ``Coulomb integrals in Liouville theory and Liouville gravity,''
  JETP Lett.\  {\bf 84}, 531 (2007).
	
\bibitem{FL2} 
  V.~A.~Fateev and A.~V.~Litvinov,
  ``Multipoint correlation functions in Liouville field theory and minimal Liouville gravity,''
  Theor.\ Math.\ Phys.\  {\bf 154}, 454 (2008)
  [arXiv:0707.1664 [hep-th]].

\bibitem{Fateev} V. Fateev, privated communication.

\bibitem{Giribet} 
  G.~Giribet,
  ``The String theory on AdS(3) as a marginal deformation of a linear dilaton background,''
  Nucl.\ Phys.\ B {\bf 737}, 209 (2006)
  [hep-th/0511252].



\bibitem{GL} 
  G.~Giribet and M.~Leoni,
  ``A Twisted FZZ-like dual for the 2D black hole,''
  Rept.\ Math.\ Phys.\  {\bf 61}, 151 (2008)
  [arXiv:0706.0036 [hep-th]].
	



\bibitem{Giribet2} 
  G.~Giribet,
  ``One-loop amplitudes of winding strings in AdS$_3$ and the Coulomb gas approach,''
  Phys.\ Rev.\ D {\bf 93}, no. 6, 064037 (2016)
  [arXiv:1511.04017 [hep-th]].


\bibitem{Stoyanovsky} 
  A.~V.~Stoyanovsky,
  ``A relation between the knizhnik-zamolodchikov and belavin-polyakov-zamolodchikov systems of partial differential equations,''
  [arXiv:math-ph/0012013].

\bibitem{RT} 
  S.~Ribault and J.~Teschner,
  ``H+(3)-WZNW correlators from Liouville theory,''
  JHEP {\bf 0506}, 014 (2005)
  [hep-th/0502048].
		
\bibitem{HS2} 
	 Y.~Hikida and V.~Schomerus,
  ``H+(3) WZNW model from Liouville field theory,''
  JHEP {\bf 0710}, 064 (2007)
  [arXiv:0706.1030 [hep-th]].

\bibitem{Ribault} 
  S.~Ribault,
  ``Knizhnik-Zamolodchikov equations and spectral flow in AdS(3) string theory,''
  JHEP {\bf 0509}, 045 (2005)
  [hep-th/0507114].

\bibitem{elGL} 
  M.~Goulian and M.~Li,
  ``Correlation functions in Liouville theory,''
  Phys.\ Rev.\ Lett.\  {\bf 66}, 2051 (1991).

\bibitem{Zamolodchikov} 
  A.~B.~Zamolodchikov,
  ``Perturbed conformal field theory on fluctuating sphere,''
  hep-th/0508044.

\bibitem{MO} 
  J.~M.~Maldacena and H.~Ooguri,
  ``Strings in AdS(3) and SL(2,R) WZW model 1.: The Spectrum,''
  J.\ Math.\ Phys.\  {\bf 42}, 2929 (2001)
  [hep-th/0001053].

		
		
\bibitem{BB}
  K.~Becker and M.~Becker,
  ``Interactions in the SL(2,IR) / U(1) black hole background,''
  Nucl.\ Phys.\ B {\bf 418}, 206 (1994)
  [hep-th/9310046].
	

\bibitem{BK} 
  M.~Bershadsky and D.~Kutasov,
  ``Comment on gauged WZW theory,''
  Phys.\ Lett.\ B {\bf 266}, 345 (1991).

\bibitem{DVV} 
  R.~Dijkgraaf, H.~L.~Verlinde and E.~P.~Verlinde,
  ``String propagation in a black hole geometry,''
  Nucl.\ Phys.\ B {\bf 371}, 269 (1992).


	

\bibitem{Lopez} 
  G.~E.~Giribet and D.~E.~Lopez-Fogliani,
  ``Remarks on free field realization of SL(2,R)(k)/U(1) x U(1) WZNW model,''
  JHEP {\bf 0406}, 026 (2004)
  [hep-th/0404231].


			









		

		
	

	

	
	
	

	
\end{thebibliography}
\end{document}